\documentstyle[11pt,paspconf,epsfig]{article}

\begin{document}
\markboth{Brotherton et al.}{FIRST 0840+3633}
\def\arcdeg{\hbox{$^\circ$}}

\title{Spectropolarimetry of FIRST 0840+3633\altaffilmark{1}}
\author{M. S. Brotherton, Wil van Breugel}
\affil{Institute of Geophysics and Planetary Physics, Lawrence Livermore National Laboratory, 7000 East Avenue, P.O. Box 808, L413, Livermore, CA 94550; mbrother@igpp.llnl.gov, wil@igpp.llnl.gov}

\author{Arjun Dey}
\affil{KPNO/NOAO\altaffilmark{2}, 950 N. Cherry Avenue, P.O. Box 26732, Tuscon,
AZ 85726; dey@noao.edu}

\author{Robert Antonucci}
\affil{Physics Department, University of California at Santa Barbara, Santa Barbara, CA 93106; ski@ginger.physics.ucsb.edu}

\altaffiltext{1}{Based on observations at the W. M. Keck Observatory.}
\altaffiltext{2}{The National Optical Observatories are operated by the
Association of Universities for Research in Astronomy under cooperative
agreement with the National Science Foundation.}

\begin{abstract}
We present Keck spectropolarimetry of a rare ``Iron Lo-BALQSO,'' 
FIRST 0840+3633.  The continuum is $\sim$4\% polarized near 2000\AA\ rest-frame,
but falls to $\sim$2\% at longer wavelengths, and maintains a relatively 
constant position angle of 50\arcdeg.  The emission lines are unpolarized.
The polarization increases up to $\sim$8\% in the low-ionization absorption
troughs of Mg II $\lambda$2800 and Al III $\lambda$1860.
The polarization and its position angle vary in a complicated manner
across the metastable Fe II absorption lines, suggesting that more than one
mechanism is at work or that the system geometry is complex.

\end{abstract}

\keywords{quasars:absorption lines, quasars:emission lines,
quasars:general, quasars:individual(FIRST 0840+3633), polarization}

\section{Introduction}

Becker et al. (1997) reported the discovery of two low-ionization BALQSOs,
radio-moderate FIRST 0840+3633 and radio-loud FIRST 1556+3517, 
found in programs to obtain optical spectra of radio-selected
QSO candidates from the VLA FIRST Survey (Becker et al. 1995).
Both BALQSOs exhibit narrow absorption lines from metastable excited
levels of Fe II and Fe III like Q0059$-$2735 (Hazard et al. 1987), 
the prototype of this extremely rare class (the ``iron Lo-BALQSOs'').

BALQSOs can be highly polarized, although the origin of the polarization 
is still being debated (see other contributions from these proceedings, e.g.,
those by Ogle, Blandford, Hines, Wills, Goodrich, Schmidt, and Cohen).
The combination of radio-selected BALQSOs (for which a radio jet orientation 
can be obtained) with high polarization (for which a polarization position
angle can be obtained) can test models in which BALQSOs are seen along
a line of sight skimming the edge of a dusty torus and polarized light
is seen along a scattered line of sight above the torus.
We report here for the first time the spectropolarimetric properties of 
an iron Lo-BALQSO, FIRST 0840+3633.

\section{Observations}

On 1996 December 10 (UT),
we observed FIRST 0840+3633 with the Low Resolution Imaging Spectrometer (LRIS;
Oke et al. 1995) in spectropolarimetry mode (Good\-rich, Cohen, \& Putney 1995)
on the 10 meter Keck II telescope.  
We used a 300 line mm$^{-1}$ grating blazed at 5000 \AA,
that, with the 1 $\arcsec$ slit (at the parallactic angle),
resulted in an effective resolution of 10 \AA\ (FWHM
of comparison lamp lines); the dispersion was 2.5 \AA\ pixel$^{-1}$.
The seeing was marginally subarcsecond.
The observation was broken into four 5 minute exposures, one for each
waveplate position (0$\arcdeg$, 45$\arcdeg$, 22.5$\arcdeg$, 67.5$\arcdeg$).
Although we observed our calibration standards with and without 
an order-blocking filter, we did not observe FIRST 0840+3633 with such
a filter so the red end of the spectrum is contaminated at a low level.

We used standard data reduction techniques inside the IRAF NOAO package,
and followed the procedures of Miller, Robinson, \& Goodrich (1988) for
calculating Stokes parameters and errors.  Figure 1 shows the 
total flux, polarization level (no debiasing scheme has been used),
polarization position angle, and polarized flux spectra.

\begin{figure}
\plotone{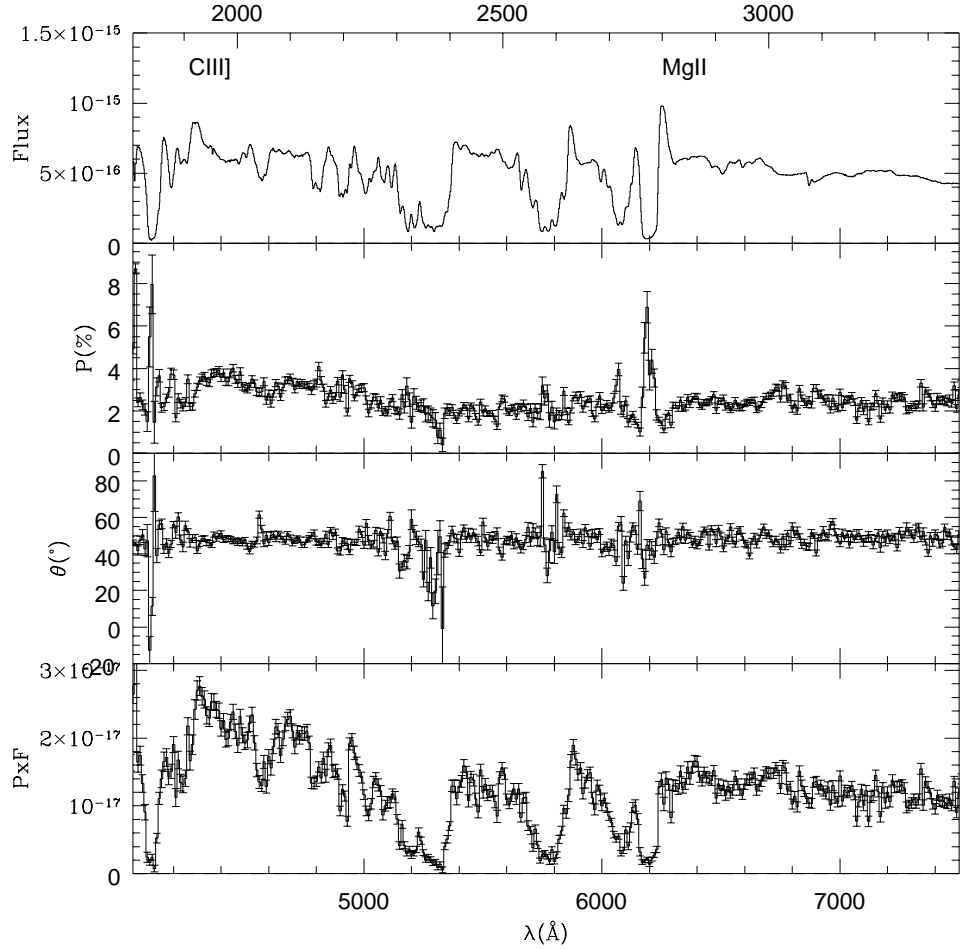}
\caption{Spectropolarimetric results for FIRST 0840+3633.  The top abscissa 
shows rest-frame wavelengths, while the bottom abscissa shows observed-frame 
wavelengths, both in \AA.  The top panel is the total flux spectrum (in
ergs s$^{-1}$ cm$^{-2}$ \AA$^{-1}$), and the C III] $\lambda$1909 and Mg II
$\lambda$2800 emission lines are labeled.  See Becker et al. 1997 (these 
proceedings) for the absorption-line identifications. The second panel from
the top shows the (biased) degree of polarization.
The third panel is the polarization position angle in degrees.
The bottom panel shows the polarized flux, the product of the top two panels.  
Error bars are 1 $\sigma$.} \label{fig-1}
\end{figure}

\section{Results}

FIRST 0840+3633 is a highly polarized BALQSO and shares many of the 
characteristics of previously studied BALQSOs (e.g., Glenn et al. 1994;
Goodrich \& Miller 1995; Cohen et al. 1995; Hines \& Wills 1995).  
These include: a significantly polarized continuum with the polarization 
increasing toward shorter wavelengths (from 2\% to 4\%), unpolarized emission 
lines (although C III] $\lambda$1909 appears to be somewhat polarized), 
and increased polarization in the low-ionization broad absorption-line 
troughs (up to 8\%). The polarization position angle is $\sim$50\arcdeg\ for
the continuum, although there is rotation evident in the absorption troughs.

The polarization structure is complex across the blended narrow
absorption-line troughs that include lines of metastable Fe II and Fe III 
as well as lines from other species.  High-resolution spectroscopy is
required to sort out just what contributes where, and in what proportions.

\section{Interpretations}

Scattering by either dust or electrons is the preferred polarization
mechanism for the continuum in BALQSOs.  BALQSOs then appear highly 
polarized because they have a favorable scattering geometry and direct, 
diluting light is more attenuated than in normal ``unpolarized'' QSOs.
There may also be some contribution 
to the polarization from resonance scattering in the emission lines
(see Ogle and Blandford's contributions), and this mechanism may explain
the polarization of C III] and the position angle rotation in the troughs in
FIRST 0840+3633.  The rise to the
blue in the continuum polarization might be the signature of dust 
scattering, but more likely it represents the dilution by unpolarized emission
from Fe II blends (the so-called ``little blue bump'').

Wampler et al. (1995) present and analyze a high-resolution spectrum 
of Q0059$-$2735.  In that object a number of individual broad and narrow-line
clouds can be identified.  Wampler et al. conclude that the low-ionization
condensations occult different parts of the background emission regions.
Thus it is probably not surprising to see complex changes and rotations
in the metastable absorption troughs of FIRST 0840+3633; 
the scattered and/or diluting light geometry may be different at these
wavelengths.

\acknowledgments

We thank the Keck Observatory staff for their assistance, and Bob Becker and 
his FIRST Survey collaborators for identifying this interesting QSO.
The W. M. Keck Observatory is a scientific partnership between the
University of California and the California Institute of Technology,
made possible by the generous gift of the W. M. Keck Foundation.
This work has been performed under the auspices of the U.S. Department of Energy
by Lawrence Livermore National Laboratory under Contract W-7405-ENG-48.

%

\begin{question}{Hines}
Could someone comment on the polarimetry of Q0059$-$2735
in comparison to FIRST 0840+3633?
\end{question}

\begin{answer}{Ogle}
Q0059$-$2735 shows complex polarization structure 
across the metastable Fe II troughs similar to what is seen in FIRST 0840+3633.
\end{answer}

\end{document}